\documentclass[conference]{IEEEtran}
\usepackage[pdftex]{graphicx}
\usepackage[keeplastbox]{flushend}
\usepackage{subcaption}
\IEEEoverridecommandlockouts

\IEEEpubid{%
   \makebox[\columnwidth]{ISBN 978-3-901882-94-4~\copyright~2017 IFIP \hfill}%
   \hspace{\columnsep}%
   \makebox[\columnwidth]{ }%
}

\begin{document}

\title{Edge-ICN and its application to the Internet of Things}

\author{
\IEEEauthorblockN{Nikos Fotiou, Vasilios A. Siris,\\ George Xylomenos, George C. Polyzos}
\IEEEauthorblockA{Mobile Multimedia Laboratory, Department of Informatics\\
School of Information Sciences and Technology\\
Athens University of Economics and Business\\
47A Evelpidon, 113 62 Athens, Greece\\
Email:\{fotiou,vsiris,xgeorge,polyzos\}@aueb.gr}
\and
\IEEEauthorblockN{Konstantinos V. Katsaros, George Petropoulos}
\IEEEauthorblockA{
Intracom SA Telecom Solutions\\
Peania, 19002, Greece\\
Email:\{konkat, geopet\}@intracom-telecom.com}
}
\maketitle

\begin{abstract}
While research on Information-Centric Networking (ICN) flourishes, its adoption seems to be an elusive goal. In this paper we propose Edge-ICN: a novel approach for deploying ICN in a single large network, such as the network of an Internet Service Provider. Although Edge-ICN requires nothing beyond an SDN-based network supporting the OpenFlow protocol, with ICN-aware nodes only at the edges of the network, it still offers the same benefits as a clean-slate ICN architecture but without the deployment hassles. Moreover, by proxying legacy traffic and transparently forwarding it through the Edge-ICN nodes, all existing applications can operate smoothly, while offering significant advantages to applications such as native support for scalable anycast, multicast, and multi-source forwarding. In this context, we show how the proposed functionality at the edge of the network can specifically benefit CoAP-based IoT applications. Our measurements show that Edge-ICN induces on average the same control plane overhead for name resolution as a centralized approach, while also enabling IoT applications to build on anycast, multicast, and multi-source forwarding primitives. 
\end{abstract}

\IEEEpeerreviewmaketitle

\section{Introduction}
\emph{Information-Centric Networking}~(ICN) has been in the spotlight of many recent research efforts around the world~\cite{Xyl2014}. However, although ICN promises to solve many problems of the current Internetworking architecture (see for example~\cite{Tro2010}) its adoption seems to be an elusive goal. The main obstacle to ICN adoption seems to be the requirement for radical changes to the entire network stack, from layer 3 technologies all the way to the application layer. In order to mitigate this problem, a number of research efforts have sprung up, investigating the potential of ICN deployments at smaller scales, as well as the possibility for backward compatibility with legacy protocols. Notable examples
of such efforts are the POINT project~\cite{Tro2016} and, more recently, Cisco's \emph{hybrid ICN}~(hICN)~\cite{Car2016} ; this paper is a further step in this direction.

The goal of the POINT project is to allow standard IP traffic to be transported over an ICN core network in a more efficient way. To achieve this, the POINT architecture provides a number of ``handlers'' implemented at \emph{Network Attachment Points}~(NAPs). These handlers perform translations between existing IP-based protocols (e.g., HTTP, CoAP, basic IP) and appropriately named objects within the ICN deployment. Similarly, the hICN project aims at allowing existing applications to operate over the named-data networking (NDN) ICN architecture~\cite{Zha2014}. In order to achieve this goal, hICN offers a scheme that allows NDN traffic to be encapsulated in IPv4/IPv6 packets, relying on a small set of ``enhanced'' routers to process these packets; legacy routers treat them as any other IPv4/IPv6 packet.  

In this work, we leverage \emph{Software-Defined Networking}~(SDN) technology to build an efficient and easily deployable ICN architecture. Adopting the decentralized approach of the fog computing paradigm, our architecture requires deploying ICN-aware nodes only at the \emph{edges} of a network, hence we refer to it as \textit{Edge-ICN}. Between these ICN-aware nodes, our architecture includes SDN switches supporting the OpenFlow protocol. Beyond the ICN-aware nodes, Edge-ICN considers legacy end-user devices and applications that are oblivious to the ICN functionality. The ICN-aware nodes (henceforth referred to as Edge-ICN nodes) translate legacy ``names'' (e.g., IP addresses,
HTTP/CoAP URIs) into ICN identifiers. An Edge-ICN node may \emph{advertise} an ICN identifier to the rest of the Edge-ICN nodes. Similarly, Edge-ICN nodes may \emph{search} for, \emph{subscribe} to, or \emph{pull} a content item based on its identifier. As a result, devices at the edge of the network, including thin end-user devices, IoT sensors/actuators and existing IoT gateways, are supported by a rich set of ICN-enabled forwarding primitives so as to engage in IoT communications e.g., request data from a semantically related set of IoT devices. Edge-ICN offers multicast and anycast capabilities that can be exploited for both requests and responses. Moreover, Edge-ICN supports content replication in multiple nodes and it creates opportunities for novel mobility, security, and service composition solutions. The SDN controller in the Edge-ICN architecture is oblivious to ICN, only being aware (as usual) of the network topology, including Edge-ICN node \emph{identifiers}, so as to be able to calculate paths between them. 

Edge-ICN borrows many concepts from the POINT architecture, but it also has key differences from it. The POINT architecture considers a centralized entity, the rendezvous point, which is aware of all available information items and their location. This information is used for computing paths between network endpoints. In Edge-ICN, information related to content availability is stored at edge nodes, with the path computation element (i.e., the SDN controller) oblivious to this information. Therefore, Edge-ICN performs no ICN-specific operation beyond the edge nodes, nevertheless, it achieves the same advantages. This separation allows for ``special purpose'' Edge-ICN nodes (e.g., nodes tailored to perform information lookup based on the constraints that an IoT protocol imposes), as well as for better traffic engineering (e.g., multiple communication paradigms). Similarly, although hICN shares the same goals as Edge-ICN (and POINT) it follows a different approach: it considers enhanced in-network devices that are aware of the ICN protocols and manipulate
packets accordingly. In Edge-ICN, apart from the Edge-ICN nodes, no other device is aware of the ICN functionality.

The structure of the remainder of this paper is as follows. In Section~\ref{sec:over} we detail our architecture, focusing on the ICN functionality of Edge-ICN nodes. In Section~\ref{sec:map} we discuss the mappings of legacy protocols to ICN and focus on the case of CoAP. In Section~\ref{sec:imp} we present our implementation and we discuss performance issues. Finally, in Section~\ref{sec:con} we provide our conclusions and plans for future work. 

\section{System overview}
\label{sec:over}

\begin{figure}
\centering
\includegraphics[width=0.90\linewidth]{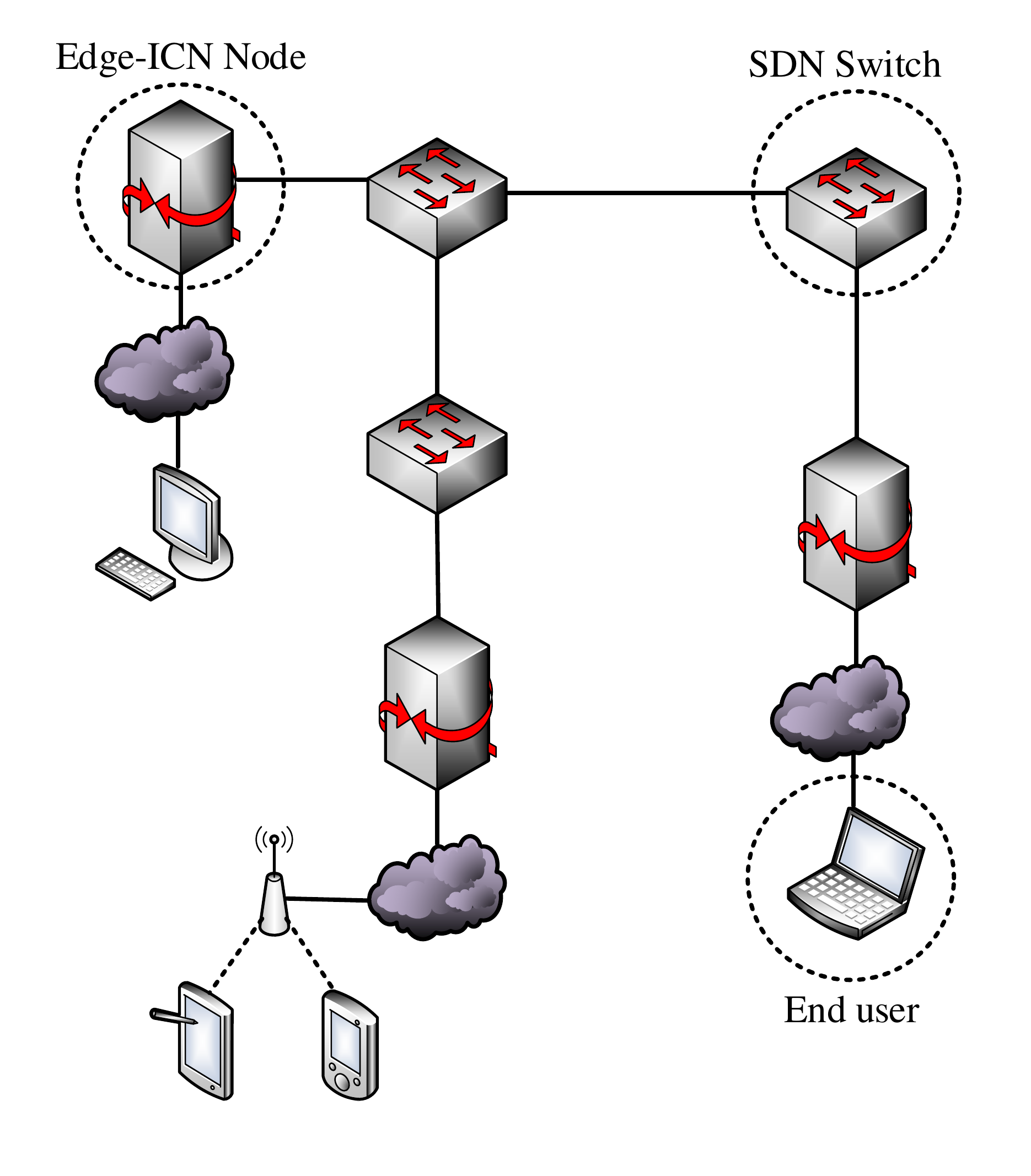}
\caption{Edge-ICN components}
\label{fig:over}
\end{figure}

Figure~\ref{fig:over} illustrates an instance of an Edge-ICN network. The main component of Edge-ICN is the \emph{Edge-ICN node}. Edge-ICN nodes act as network attachment points for legacy end-user devices and are interconnected through a network composed of SDN switches. Edge-ICN nodes do not disrupt existing protocols and services, but enhance the network with ICN benefits. All Edge-ICN nodes are identified by \emph{at least} one identifier denoted by $Node_{ID}$. These identifiers are assumed to be distributed on network bootstrapping, and managed by the SDN controller. The semantics of $Node_{ID}$ are deployment strategy specific. The only requirement imposed by our architecture 
is that a $Node_{ID}$ should fit in the payload of single IPv6/UDP packet. Moreover, our architecture considers a virtual node identifier, denoted by $all.nodes$, whose semantics are ``all Edge-ICN nodes". Depending on the deployment strategy, other virtual node identifiers can also be considered, e.g., $all.coap.nodes$, i.e., all Edge-ICN nodes that support CoAP handling, $all.http.nodes$, i.e., all Edge-ICN nodes that support HTTP handling, and so on.

For all traffic between Edge-ICN nodes, Bloom filter based forwarding is used~\cite{Ree2016}. To realize this, each link of the SDN network is assigned a unique identifier, called its $Link_{ID}$, which is a fixed-size Bloom filter-based vector~\cite{Bloom1970}. Each $Link_{ID}$ is mapped to an arbitrary bitmask with length equal to two IPv6 addresses and corresponding OpenFlow rules are installed at its adjacent switch interfaces. The forwarding identifier that realizes the communication between Edge-ICN nodes is a Bloom filter encoding the link identifiers of the path that a packet should follow, by simply ORing these identifiers. An interesting property of this type of forwarding is that by ORing the encodings of two paths $A \rightarrow B$ and $A \rightarrow C$, the encoding of a multicast path from $A$ to $B,C$ is derived. The solution in~\cite{Ree2016} encodes Bloom filters in the IPv6 source and destination fields of an IPv6 packet and uses the OpenFlow arbitrary mask match in switches to make forwarding decisions. Therefore, this forwarding technique is implemented using standard OpenFlow mechanisms and without modifying the structure of network packets; to a network monitoring tool, these packets appear to be ordinary
IPv6 packets, but with ``strange'' addresses.  An SDN controller can calculate a Bloom filter for a path between two Edge-ICN nodes, as well as between an Edge-ICN node and
$all.nodes$ (or any other virtual node identifier). 

When an Edge-ICN node wants to send a packet towards another Edge-ICN node, it has the following options: (a) it adds the $Node_{ID}$ of the destination (or a virtual node identifier) in the payload of a special new packet, and sets the IPv6 source and destination fields of the packet to pre-defined constant value, (b) it directly encodes a Bloom filter in the IPv6 source and destination fields. In order to distinguish between  these two packet types, a different value for the Ethernet type field is used. For option (b), packet switching is performed based on the aforementioned forwarding principles.

For option (a), packet forwarding requires first resolving the node identifier to a Bloom filter. The first SDN switch that receives the special packet does not have a rule to switch it, hence it forwards it to the SDN controller as an Openflow \textit{PacketIn} message~\cite{Openflow}. The SDN controller extracts the provided $Node_{ID}$ of the destination (or any virtual node identifier), and given its knowledge of the underlying SDN topology and assigned node and link identifiers calculates the appropriate Bloom filter, as well as a Bloom filter for the reverse path (if applicable), and returns this information to the switch as an Openflow \textit{PacketOut} message. Finally, the switch forwards this message to the Edge-ICN node that originally sent the packet. The Edge-ICN node can now use the provided Bloom filter for all subsequent requests destined for that $Node_{ID}$; all these requests are forwarded by the SDN network without further communicating with the controller. 

It should be noted that the initial packet contains only the destination $Node_{ID}$, hence the SDN controller learns less information about a user's preferences compared to, for example, POINT~\cite{Tro2016}, where content names are resolved rather than edge node names. In addition, the aforementioned process does not require the presence of a centralized ICN-specific  ``rendezvous entity''~\cite{Tro2016} to match advertisements between Edge-ICN nodes, hence it does not impose further deployment requirements to network administrators.

A salient feature of Edge-ICN is its \emph{anycast} capability. In an Edge-ICN network a (virtual) $Node_{ID}$ does not have to be unique, i.e., there can be multiple Edge-ICN nodes sharing the same identifier; recall here that an edge-ICN node may be associated with multiple identifiers. The first time an SDN switch encounters a packet destined for a $Node_{ID}$, it asks the SDN controller for forwarding instructions; if there exist multiple nodes sharing this identifier, the controller can implement a deployment specific anycast strategy and select the most appropriate. 

\subsection{ICN operations} 
\begin{figure*}[ht!]
    \centering
    \begin{subfigure}{0.45\textwidth}
        \centering
        \includegraphics[width=\textwidth]{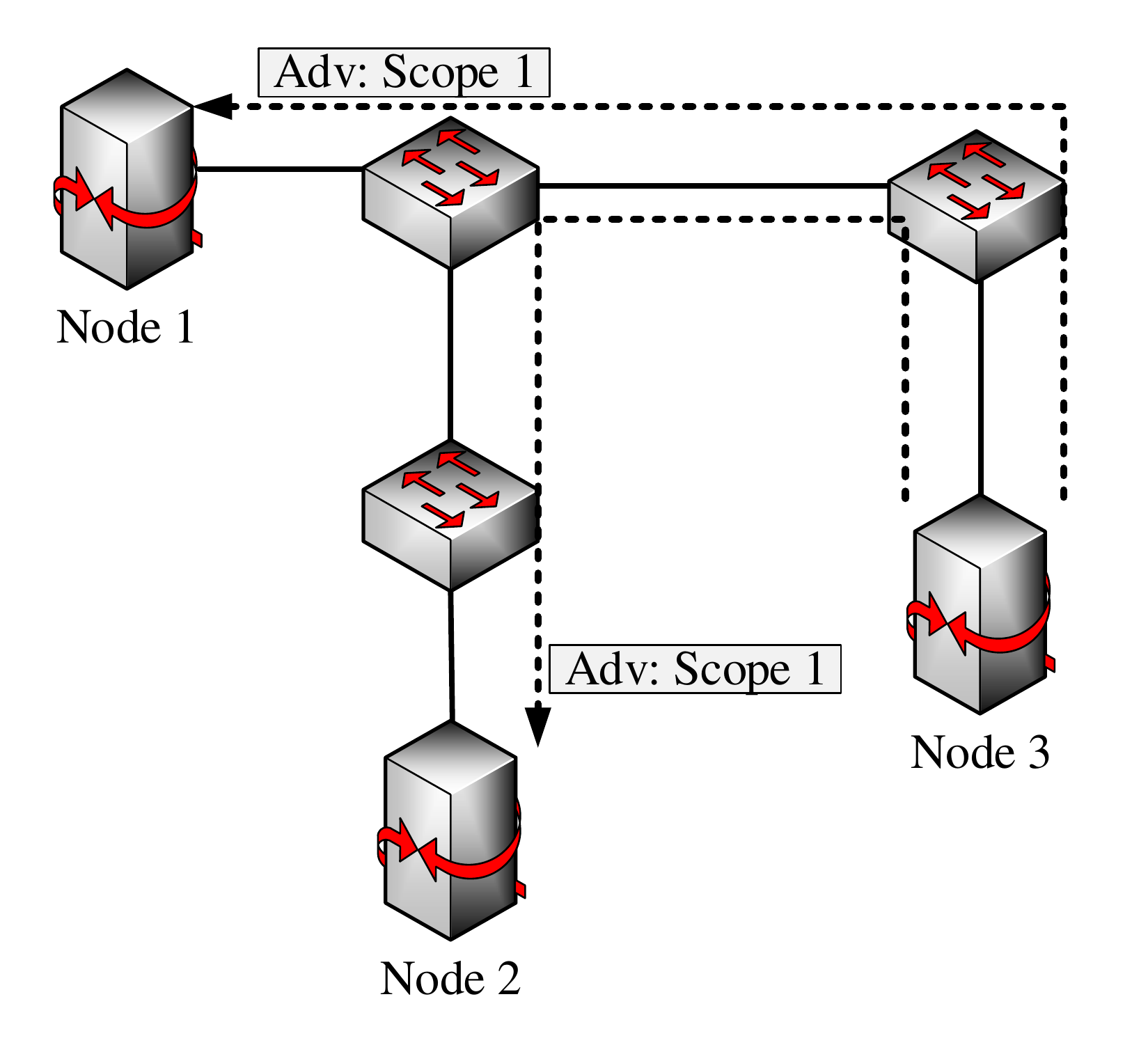}
        \caption{}
    \end{subfigure}
    ~ 
    \begin{subfigure}{0.45\textwidth}
        \centering
        \includegraphics[width=\textwidth]{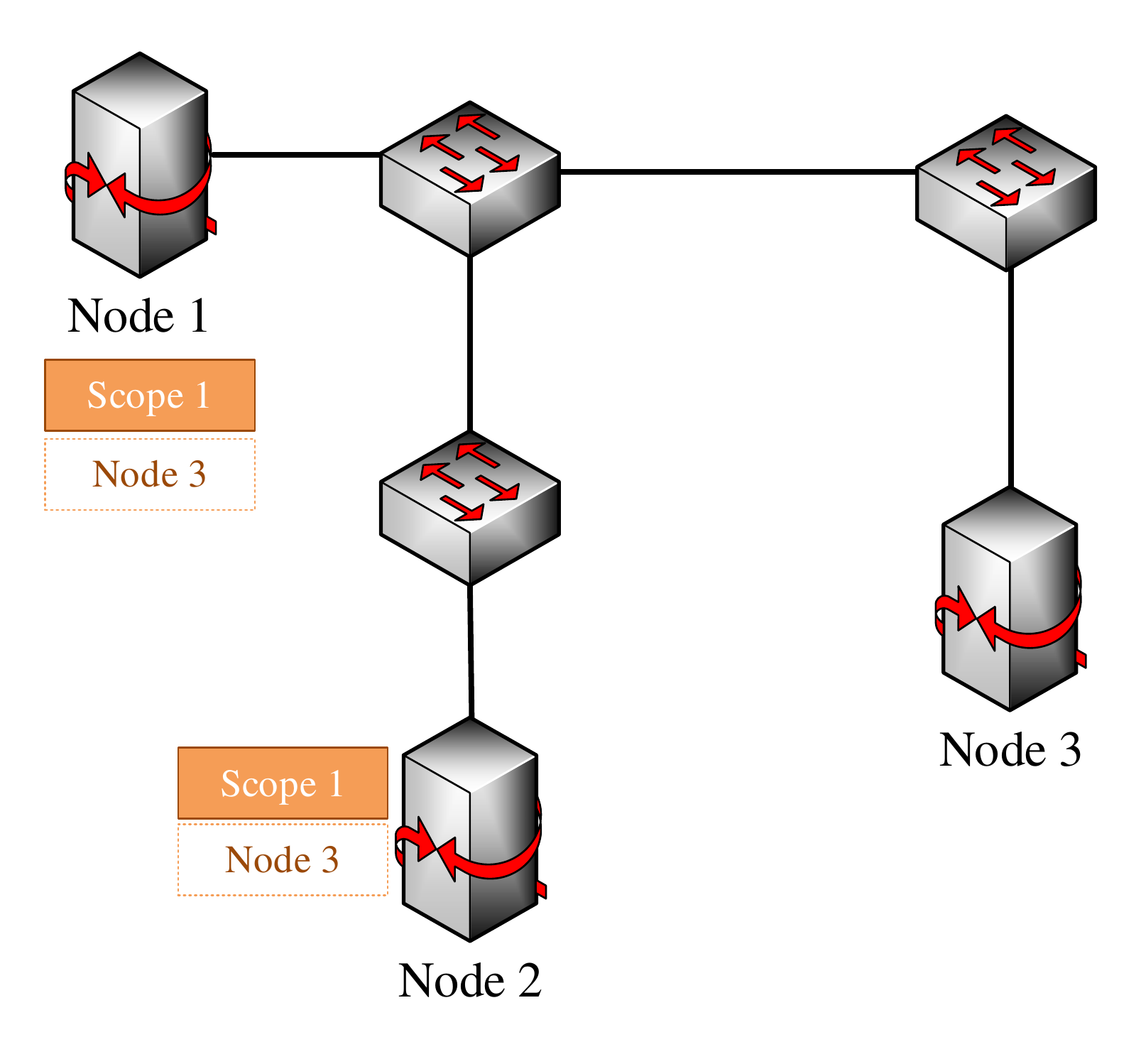}
        \caption{}
    \end{subfigure}
    \caption{Scope advertisement}
    \label{fig:adv}
\end{figure*}

Edge-ICN adopts the information organization semantics of the Publish-Subscribe Internet architecture~\cite{Xyl2012}, i.e., all information items are grouped in \emph{scopes}. In terms of legacy protocols, a scope can be an IP address, a domain name, etc. Edge-ICN nodes may \emph{advertise} a scope they ``know'' by sending an advertisement message to the $all.nodes$ identifier; depending on the deployment strategy, they could use a more fine-grained virtual node identifier. An advertisement message contains the scope identifier, the $Node_{ID}$ of the sender and, optionally, a payload. When an Edge-ICN node receives such an advertisement, it updates a lookup table that contains tuples of the form $[Scope Identifier, Node_{ID}]$. An example of scope advertisement is illustrated in Figure~\ref{fig:adv}. In this example $Node 3$ advertises a scope, i.e., $Scope 1$; the advertisement is received by both $Node 1$ and $Node 2$, which update their lookup tables accordingly. Edge-ICN supports \emph{scope replication}, therefore the same scope can be advertised by multiple nodes. In that case, lookup tables should contain all node identifiers that have advertised a specific scope. Finally, Edge-ICN supports advertisements where the sender field is set to a virtual node identifier (e.g., $all.coap.nodes$), which means that all nodes belonging to that group ``know'' this scope.

Edge-ICN nodes can \emph{subscribe} to information items stored in a scope. In contrast to~\cite{Xyl2012}, Edge-ICN nodes can not request an item stored in a scope that has not yet been created. When a node wants to send a subscription message, it uses its lookup table and selects an appropriate destination $Node_{ID}$; if there are multiple candidates, then the selection process is based on a deployment specific strategy e.g., to achieve load balancing. It is noted that such strategies can further express application semantics e.g., retrieving sensor information from selected IoT gateways, subject to the time of the day or geographical area. In this way, application logic gets further integrated within the edge network. As a next step, the node constructs a network packet that includes the following: the identifier of the desired item, the identifier of the scope in which this item belongs to, the subscribing node identifier and, optionally, a payload (which can contain the Bloom filter of the reverse path). This packet is forwarded to the selected $Node_{ID}$, using the process defined in the previous subsection.

A feature of Edge-ICN that is of particular importance for ICN applications, is its ability to forward a content subscription to multiple Edge-ICN nodes. This feature, which can be considered as a generalization of the anycast capability, is implemented by ``grouping'' target nodes under the same (virtual) $Node_{ID}$; when an SDN controller receives a request for such a  $Node_{ID}$ it responds with a Bloom filter that covers all nodes of the group. Note here the difference between multiple nodes advertising the same scope, which lets the Edge-ICN node select where to send a subscription, and using a virtual $Node_{ID}$, where the network returns all members of the corresponding group. This feature is particularly useful for information searching, as well as for ``pulling'' information items. With Edge-ICN we can choose to either advertise fine-grained scopes or use ``generic'' scopes associated with virtual node identifiers. In the former case, we essentially broadcast specific advertisements, corresponding to a push model; in the latter case, we broadcast the content requests, corresponding to a pull model. The choice should take into account the expected popularity of the content and its lifetime, or its mobility in case of mobile content, in addition to the cost of broadcasting advertisements to all edge nodes, which depends on the number of edge nodes and the network topology.
  
The pull model supported by Edge-ICN facilitates the deployment of privacy-enhancing solutions. For instance, a node may encrypt the desired content identifier and subscription payload using Identity-Based Encryption (as for example described in~\cite{Fot2016b}) and broadcast the subscription message to all nodes belonging to a virtual nodes identifier; only the node that holds the desired item should be able to decrypt the encrypted parts of the message. 

\section{Mapping legacy protocols to ICN}
\label{sec:map}

ICN has been regarded as a promising candidate for building IoT architectures. In fact, RFC 7476~\cite{rfc7476} specifies IoT as a baseline ICN scenario. To this end, many ICN-based IoT architectures has been proposed (e.g.,~\cite{Bis2013}, \cite{Gri2014}, \cite{Pol2015}). However, all these architectures follow a clean-slate approach. The POINT project specifies an approach for supporting IP-enabled IoT devices using the standard CoAP protocol; Edge-ICN builds on this approach. Specifically, Edge-ICN uses the solutions developed by the POINT project in order to map legacy protocol name spaces to ICN identifiers~\cite{Tro2016}. When it comes to the IoT and the CoAP protocol, the approach described in~\cite{Fot2016} is followed. 

CoAP~\cite{rfc7252} is a lightweight protocol, designed to be the ``HTTP of the IoT." The CoAP interaction model is similar to the client/server model of HTTP: a CoAP client requests a resource from a server; if the resource is available, the server responds, otherwise it simply ACKnowledges the request and responds asynchronously when the resource becomes available. CoAP resources are identified  by a URI, similar to HTTP URIs: the host part of this URI is mapped to an ICN scope. Edge-ICN nodes that are aware of a CoAP resource (e.g., through static configuration or by using a discovery protocol such as CoAP Observe~\cite{rfc7641}), advertise the appropriate scope identifier using the procedures described in the previous section. In order for a CoAP client to take advantage of Edge-ICN, it should be configured to use an Edge-ICN node as a CoAP proxy. When an Edge-ICN node 
receives a CoAP request (from the IP world), it extracts the \emph{URI-host} and checks, using its lookup table, if there is a scope identified by the same name (i.e., the URI-host). If there is no such scope, it means that the requested resource is not accessible through ICN, so the Edge-ICN node process the packet following the standard CoAP procedures; otherwise, it creates a subscription message that contains in its payload the CoAP request headers and (if applicable) a reverse path Bloom filter.

As discussed in~\cite{Fot2016} significant gains can be achieved by aggregating requests and by multicasting responses (by ORing the reverse path Bloom filters). This feature is particularly useful when (CoAP) responses are sent asynchronously or when the CoAP Observe extension is used. Another benefit of ICN to CoAP, also  reported in~\cite{Fot2016}, is that due to ICN's native support for multicast it is easier to support CoAP Group Communication~\cite{rfc7390}, which allows a CoAP client to send a CoAP request to multiple CoAP servers simultaneously. In an IP network, CoAP group communication is implemented using IP multicast. However, by forwarding CoAP requests over an ICN network, CoAP endpoints do not have to support IP multicast, neither do they have to maintain excessive state.\footnote{For more details, interested readers are referred to ~\cite{Fot2016}.} 

The CoAP Group Communication RFC also defines the notion of ``context-based'' URI-hosts, e.g., $all.west.building6$, which should be mapped to an IP multicast address that corresponds to all CoAP endpoints  located in the west wing of building 6. In order to implement this behavior in an IP network (a) DNS servers should be modified, and (b) all CoAP servers (or their gateways) should join a priori \emph{all} possible IP multicast groups. With Edge-ICN and the notion of the virtual node identifier this functionality can be implemented in a much easier way. SDN controllers can be configured with virtual node identifiers, such as ``west'', ``building6'', ``temperature'', as well as with the mappings from this identifiers to real $Node_{ID}s$; when a Bloom filter is requested, for example, to $all.west.building6$, the SDN controller can find the $Node_{ID}s$ that belong to all three virtual node identifiers and construct the appropriate Bloom filter. 

An advantage of distributing the scope lookup process to the edge nodes, as opposed to having a centralized entity as in~\cite{Tro2016}, is that application specific Edge-ICN nodes can be considered. For example, in an IoT architecture there can be ``enhanced'' Edge-ICN nodes that support some form of service description in scope advertisements (e.g., using the CoRE link format~\cite{rfc6690}), allowing easier
service composition and service chaining, or even M2M communication. Similarly, the state stored in a CoAP resource directory may as well be distributed to specialized
Edge-ICN nodes supporting ``smarter'' subscriptions (e.g., ``send a notification if the temperature falls below X degrees''). 

\label{sec:imp}
\begin{figure}
\centering
\includegraphics[width=0.90\linewidth]{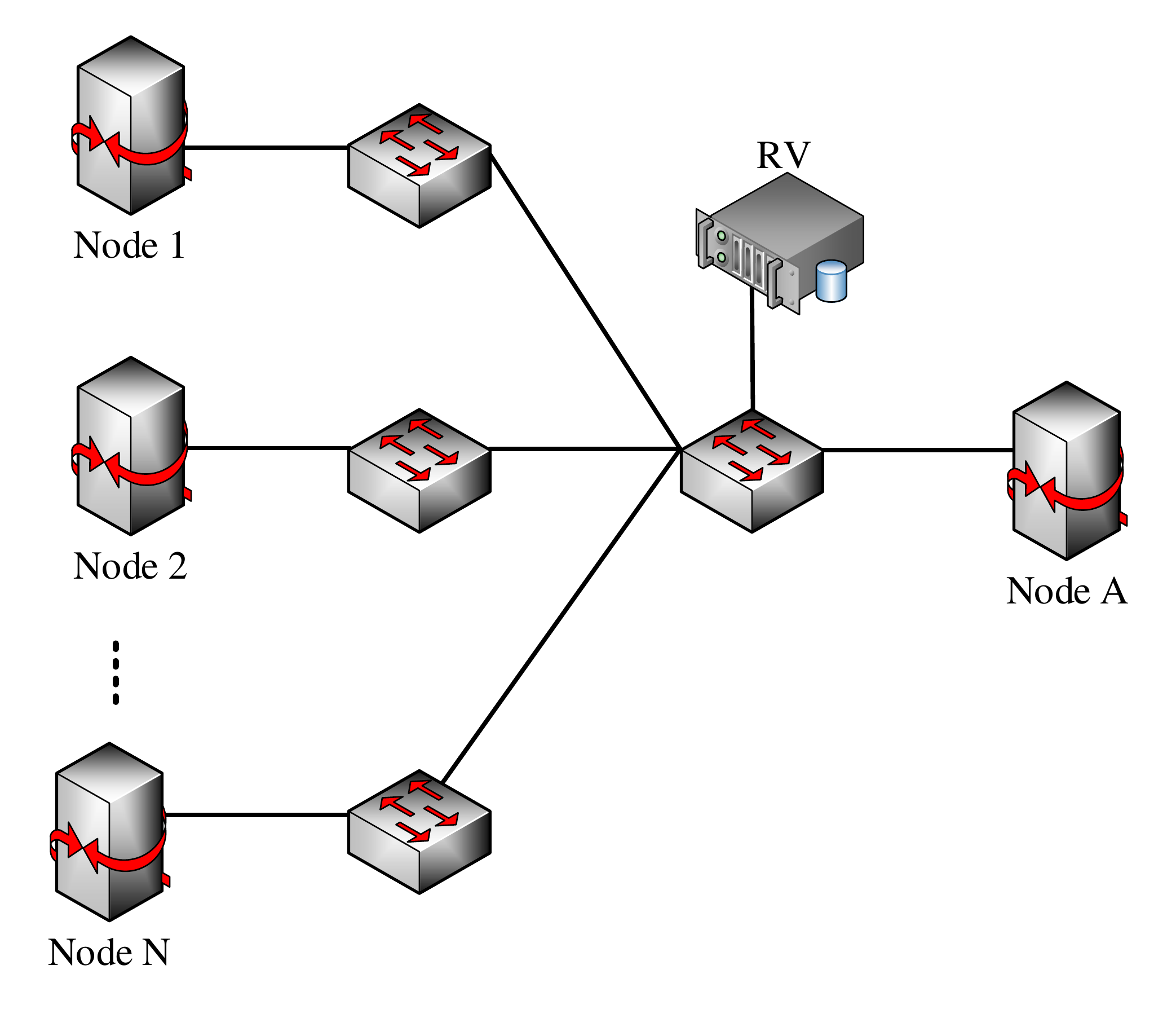}
\caption{Evaluation topology. RV stands for POINT's rendezvous point.}
\label{fig:eval}
\end{figure}

\begin{figure*}
    \centering
    \begin{subfigure}{0.3\textwidth}
        \centering
        \includegraphics[width=\textwidth]{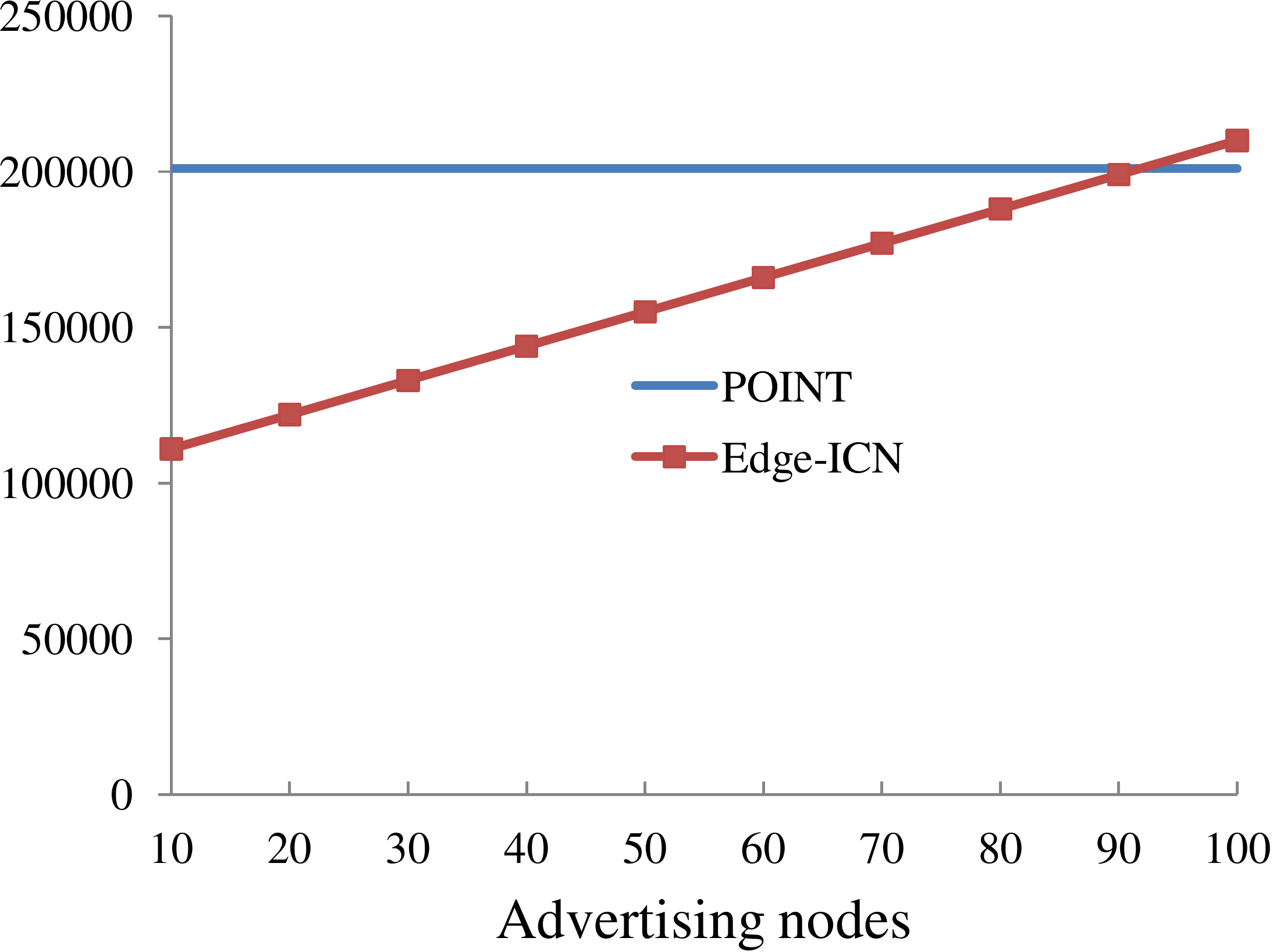}
        \caption{100 subscribing nodes and 1000  scopes.}
    \end{subfigure}
    ~ 
     \begin{subfigure}{0.3\textwidth}
        \centering
        \includegraphics[width=\textwidth]{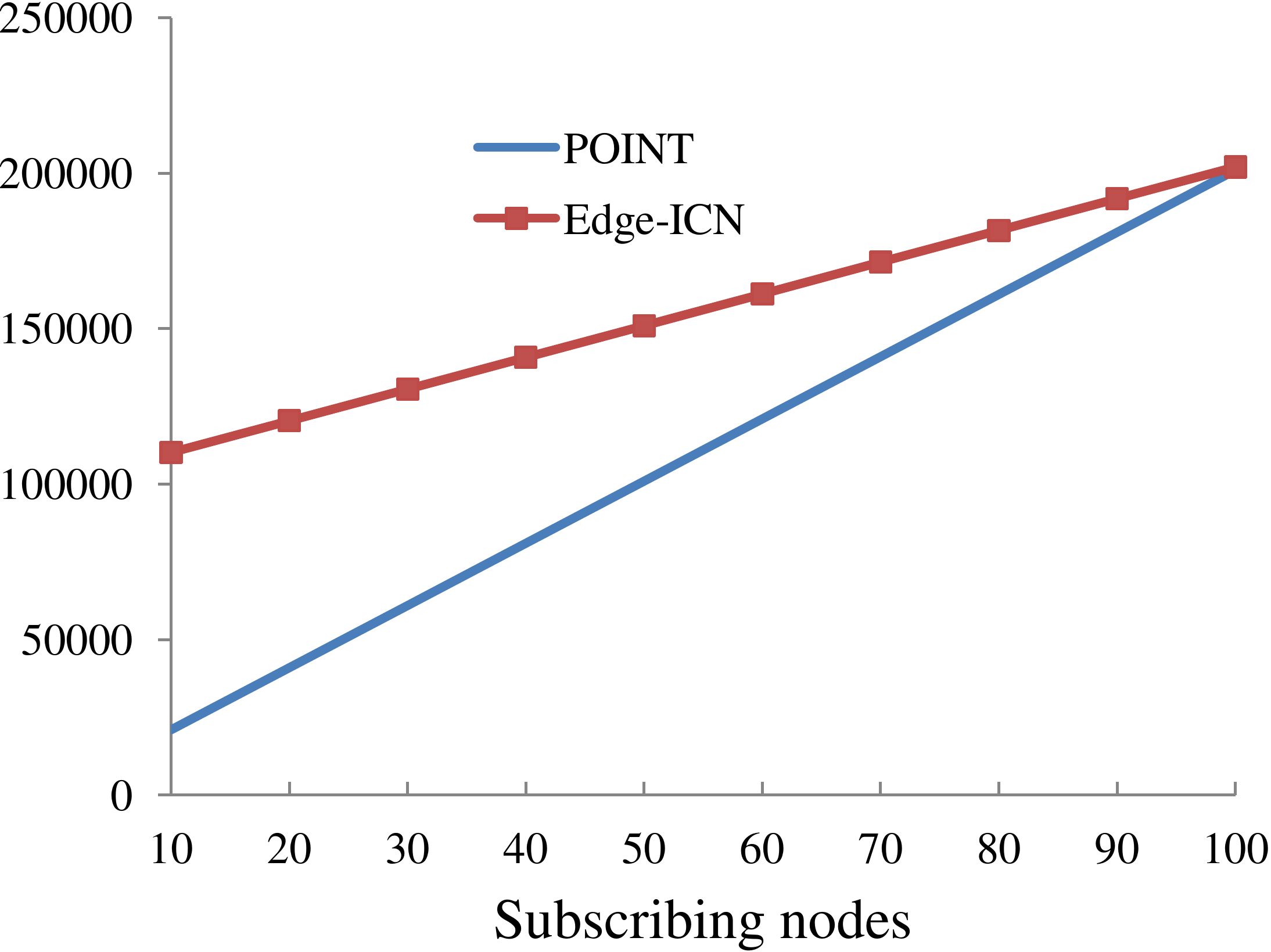}
        \caption{100 advertising nodes and 1000 scopes.}
    \end{subfigure}
    ~
     \begin{subfigure}{0.3\textwidth}
        \centering
        \includegraphics[width=\textwidth]{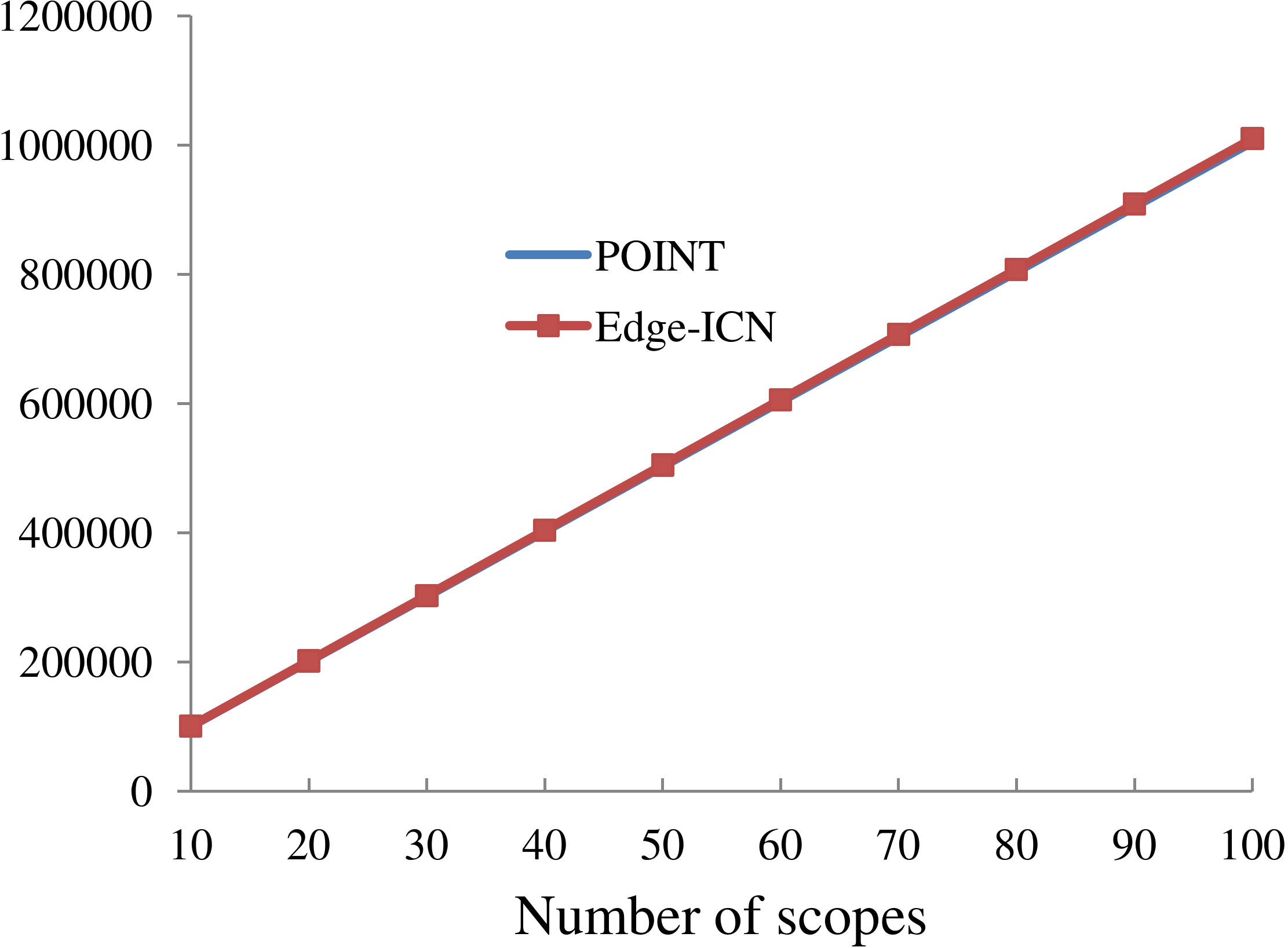}
        \caption{100 advertising nodes and 100 subscribing nodes.}
    \end{subfigure}
    \caption{Communication overhead measured in times $l$.}
    \label{fig:res}
\end{figure*}

\section{Implementation and Evaluation}
To validate the feasibility of our concept, we have implemented
Edge-ICN using the mininet network emulator~\cite{Lan2010}, Open vSwitch~\cite{Pfa2015}, and the
POX SDN controller~\cite{Gud2008}.\footnote{Source code is available at https://github.com/mmlab/edge-icn.}

A concern that may arise is related to the communication overhead introduced due to scope advertisements to $all.nodes$ (or to other virtual node identifiers). At first glance it may seem that storing all advertisements in a centrally located entity e.g., as in POINT, where all advertisements are sent to a centralized ``rendezvous point'', is more optimal. Nevertheless, this is not the case, since content subscriptions also must be be sent to that same entity. In order to illustrate this intuition, consider the topology of Figure~\ref{fig:eval}.
In this figure there exist two types of nodes: N+1 edge nodes and 1 centrally located node. The latter node is referred to as the rendezvous point (RV) and it is used by POINT to store all scope advertisements. In a nutshell, content dissemination in POINT is implemented by executing the following
steps:\footnote{For clarity reasons, the terminology has been adapted to the context of this paper.}
\begin{itemize}
\item \textbf{Step 1} A node (referred to as the server NAP, or sNAP for short) sends a scope advertisement to the RV.
\item \textbf{Step 2} Another node (referred to as the client NAP, or cNAP for short) ``advertises'' a content request to the RV.
\item \textbf{Step 3} The RV responds with a Bloom filter that encodes a path towards an sNAP.
\item \textbf{Step 4} The cNAP sends a content request to the sNAP using the provided Bloom filter.
\item \textbf{Step 5} The sNAP responds with the content item.
\end{itemize} 
For all subsequent requests by the same subscriber for items \emph{belonging to the same scope} steps 1, 2 and 3 are omitted. Similarly, the steps required to transfer a content item in Edge-ICN are the following:\footnote{It is assumed that Edge-ICN nodes have already advertised at least one item, hence they know the Bloom filter that corresponds to the virtual node identifier $all.nodes$.}
\begin{itemize}
\item \textbf{Step 1} An Edge-ICN node sends a scope advertisement to all other nodes.
\item \textbf{Step 2} Another node sends a subscription which is received by the SDN controller.
\item \textbf{Step 3} The controller responds with a Bloom filter that encodes a path towards a $Node_{ID}$.
\item \textbf{Step 4} The node sends the subscription using the provided Bloom filter.
\item \textbf{Step 5} The node that received the subscription responds with the content item.
\end{itemize} 
For all subsequent requests by the same node towards \emph{the same $Node_{ID}$} steps 1, 2 and 3 are omitted.

We now compare the communication overhead of these two approaches, by making the following assumptions: (i) all messages are of equal length (ii) all paths between any NAP and the RV in POINT are of equal length $l$, (iii) all paths between any Edge-ICN node and the SDN controller are of equal length $l$ (i.e., the same length
as in assumption (ii)) (iv) all paths between any NAP, as well as between any two Edge-ICN nodes are of equal length $2l$, (v) the length of the multicast tree used in Edge-ICN for advertising a scope to all nodes is $1+l*|nodes|$, where $|nodes|$ is the number of the nodes that receive the advertisement. We can now calculate the communication overhead of each architecture assuming the same topology. In our calculations, we do not consider steps 4 and 5 of both architectures, as they introduced the same overhead. 

We consider a number of scopes equally distributed to a  number of advertising nodes. Moreover, a number of subscribing nodes request one item per scope. In Edge-ICN an advertising node advertises a scope to all nodes (i.e., advertising + subscribing nodes). Figure~\ref{fig:res} shows the communication overhead measured in times $l$ (i.e., a packet that traverses a network path of size $l$ is measured to introduce overhead 1) as a function of the number of (a) advertising nodes, (b) subscribing nodes, and (c) scopes. It can be observed that when the number of subscribing nodes is higher than the number of the advertising nodes, Edge-ICN introduces less overhead than POINT. In contrast, when the number of advertising nodes is bigger that the number of subscribing nodes, Edge-ICN introduces more overhead than POINT. The reason is that Edge-ICN has higher costs for content advertisement but lower costs for content subscription than POINT. When the number of subscribing and advertising nodes are equal, then both architectures behave almost the same.

Of course the results are obtained through analysis. Thorough simulations with realistic topologies and workloads will be considered in follow-up work.

\section{Conclusions and future work}
\label{sec:con}
In this paper we presented Edge-ICN, an efficient and easily deployable ICN architecture. Edge-ICN aims at extending existing networks with ICN capabilities by leveraging SDN technology. In particular, Edge-ICN exploits an SDN-based substrate to build communication paradigms that can enable ICN applications, exploiting push, pull, anycast and mutlicast. By implementing all ICN logic in edge nodes and by mapping legacy protocols to ICN-functions, Edge-ICN requires no modifications to end user devices or protocols nor to the (now SDN-based) core network infrastructure. Additionally, existing applications can benefit from Edge-ICN despite the fact they are oblivious to the ICN functions of the architecture. To this end, we discussed the gains realized by a CoAP-based IoT application when Edge-ICN is used. 

The work reported in this paper is at an early stage of development and it assumes reliable control messaging, stable and legitimate nodes, and only legacy applications. It is in our
immediate plans to revisit all these assumptions. In particular, we will investigate SDN monitoring tools that are able to detect node failures and provide churn handling mechanisms, we will consider packet losses in the control plane, and we will design node mobility solutions. Moreover, we will explore the benefits that can be gained by considering ICN-aware applications in addition to legacy ones. Edge-ICN is designed to be a network-wide solution and assumes that all Edge-ICN nodes belong to the network operator. An interesting twist of this assumption is to consider user-owned Edge-ICN devices; this case will create opportunities for novel applications, as well as new challenges.

\section*{Acknowledgments}
This research was supported by the EU funded H2020 ICT project POINT, under contract 643990.
       

\end{document}